\documentclass[twocolumn, amsmath,amssymb, superscriptaddress]{revtex4}

\usepackage{dcolumn}
\usepackage{bm}
\usepackage{amsmath}
\usepackage{amsfonts}
\usepackage{graphics}
\usepackage[dvips]{graphicx}
\usepackage{times}
\usepackage{setspace}

\begin{document}

\title{Cross-Correlations between Volume Change and Price Change}
\author{Boris~Podobnik}
\affiliation{Center for Polymer Studies and Department of Physics, Boston University, Boston, Massachusetts 02215, USA}
\affiliation{Zagreb School of Economics and Management, 10000  Zagreb, Croatia}
\affiliation{Faculty of Civil Engineering, University of Rijeka, 51000 Rijeka, Croatia}
\author{Davor Horvatic}
\affiliation{Faculty of Science, University of Zagreb, 10000 Zagreb, Croatia}
\author{Alexander M. Petersen}
\affiliation{Center for Polymer Studies and Department of Physics, Boston University, Boston, Massachusetts 02215, USA}
\author{H. Eugene Stanley}
\affiliation{Center for Polymer Studies and Department of Physics, Boston University, Boston, Massachusetts 02215, USA}
\date{\today}

\begin{abstract}
In finance, one usually deals not with prices but with growth rates $R$,
defined as the difference in logarithm between two consecutive prices.
Here we consider not the trading volume, but rather the volume 
 growth rate $\tilde R$, the
difference in logarithm between two consecutive values of trading
volume. To this end, we use several methods to
analyze the properties of volume changes $|\tilde R|$, and their
relationship to price changes $|R|$.
We analyze $14,981$ daily recordings of the
S\&P 500 index over the 59-year period 1950--2009, and find power-law
{\it cross-correlations\/} between $|R|$ and $|\tilde R|$ using 
detrended cross-correlation analysis (DCCA).
We introduce
a joint stochastic process that models these cross-correlations.
Motivated by the relationship between $| R|$ and $|\tilde R|$, 
we estimate the tail exponent
${\tilde\alpha}$ of the probability  density function 
$P(|\tilde R|) \sim |\tilde R|^{-1 -\tilde\alpha}$  for both the
 S\&P 500 index
as well as the collection of 1819 constituents of the New York Stock
Exchange Composite index on 17 July 2009.
As a new  method to estimate $\tilde\alpha$, we calculate
the time intervals $\tau_q$ between events where $\tilde R>q$. We
demonstrate that $\bar\tau_q$, the average of $\tau_q$, obeys
$\bar   \tau_q \sim q^{\tilde\alpha}$. We find $\tilde\alpha
\approx 3$. Furthermore, by
aggregating all $\tau_q$ values of 28 global financial indices, we also
observe
an approximate inverse cubic law.
\end{abstract}

\maketitle

There is a saying on Wall Street that ``It takes volume to move stock
prices.''  A number of studies have analyzed the relationship between
 price changes and the trading volume in financial markets
\cite{Ying,Crou,Clark,Epps,Rogalski,Cornell,Tauc,Gram,Karp,Lamo90,Gall,tauch96,Gabaix,Gabaix06x}.
Some of these studies \cite{Ying,Clark,Epps,Rogalski,Cornell} have found
a positive relationship between price change and the trading volume. In
order to explain this relationship, Clarke assumed that the daily price
change is the sum of a random number of uncorrelated intraday price
changes \cite{Clark}, so predicted that the variance of the daily price
change is proportional to the average number of daily transactions. If
the number of transactions is proportional to the trading volume, then
the trading volume is proportional to the variance of the daily price
change.

The cumulative distribution function (cdf) of 
 the absolute  logarithmic price {\it
  change\/} $|R|$ obeys a power law
\begin{equation}
P(|R|>x)\sim  x^{-\alpha}.
\end{equation}
It is believed \cite{Lux96,Gopi98,Gopi99,Plerou99} that $\alpha\approx
3$ (``inverse cubic law''), outside the range $\alpha<2$ characterizing
a L\'evy distribution \cite{Mand63,Plerou99}.  A parallel analysis of
$Q$, the volume traded, yields a power law
\cite{Gopi00,Plerou07xx,Racz09,Plerou09reply,Gopi01,Kert05,kertesz,Farm04,Plerou04xx}
\begin{equation}
P(Q>x)\sim  x^{-\alpha_Q}.
\end{equation}
To our knowledge, the logarithmic volume {\it change}---$\tilde R$ and
its relation to the logarithmic price change $R$---has not been
analyzed, and this analysis is our focus here.

\section{Data Analyzed}

\begin{itemize}

\item[{A.}] We analyze the S\&P500 index recorded daily over the 59-year
  period January 1950 -- July 2009 (14,981 total data points).

\item[{B.}] We also analyze  1819 New York Stock Exchange (NYSE)
  Composite members comprising this index on 17 July 2009, recorded at
  one-day intervals (6,794,830 total data points). Both data sets are
  taken from http://finance.yahoo.com. Different companies comprising
  the NYSE Composite index have time series of different lengths. The
  average time series length is 3,735 data points, the shortest time
  series is 10 data points, while the longest is 11,966 data points. If
  the data display scale independence, then the same scaling law should
  hold for different time periods.

\item[{C.}] We also analyze 28 worldwide financial indices from
  http://finance.yahoo.com recorded daily.

\begin{itemize}

\item[{(i)}] 11 European indices (ATX, BEL20, CAC 40, DAX, AEX General,
  OSE All Share, MIBTel,  Madrid General,
  Stockholm General, Swiss Market, FTSE 100),

\item[{(ii)}] 12 Asian indices (All Ordinaries, Shanghai
  Composite, Hang Seng, BSE 30, Jakarta Composite, KLSE
  Composite, Nikkei 225, NZSE 50, Straits Times, Seoul Composite,
  Taiwan Weighted, TA-100),  and

\item[{(iii)}] 5 American and Latin American indices (MerVal,
  Bovespa,  S\&P TSX Composite, IPC,  S\&P500 Index).


\end{itemize}

\end{itemize}

\noindent For each of the 1819 companies and 28 indices, we calculate
over the time interval of one day the logarithmic change in price
$S(t)$,
\begin{equation}
R_{t}\equiv\ln\left(\frac{S(t + 1)}{S(t)}\right), 
\label{RS}
\end{equation}  
and also the logarithmic change in trading volume $Q(t)$ \cite{Ausloos},
\begin{equation}     
\tilde R_{t}\equiv\ln\left({Q(t+ 1)\over Q(t)}\right).
\label{RV}
\end{equation}
For each of the 3694 time series, we also calculate the absolute values
$|R_t|$ and $|\tilde R_t|$ and define the ``price volatility''
\cite{Liu99} and ``volume volatility,'' respectively,
\begin{equation}     
V_R\equiv\frac{|R_t|}{\sigma_R}
\end{equation}
and
\begin{equation}
V_{\tilde R}\equiv\frac{|\tilde R_t|}{\sigma_{\tilde R}},
\label{RVnorm}
\end{equation}
where $\sigma_R\equiv(\langle|R_t|^2 \rangle-\langle
|R_t|\rangle^2)^{1/2}$ and $\sigma_{\tilde R}\equiv(\langle|\tilde
R_t|^2 \rangle-\langle |\tilde R_t|\rangle^2)^{1/2}$ are the respective
standard deviations.
 
\section{Methods}
 
Recently, several papers have studied the return intervals $\tau$
between consecutive price fluctuations above a volatility threshold
$q$. The pdf of return intervals $P_q(\tau)$ scales with the mean return
interval $\tau$ as \cite{Yama05,Wang06,Wang08}
\begin{equation}
P_q(\tau)=\overline{\tau}^{-1}f\left({\tau\over\overline{\tau}}\right),
\end{equation}
where $f(x)$ is a stretched exponential.  Similar scaling was found for
intratrading times (case $q=0)$ in Ref.~\cite{Ivanov04}. In this paper
we analyze either (i) separate indices or (ii) aggregated data mimicking
the market as a whole. In case (i), e.g., the S\&P500 index for any $q$,
we calculate all the $\tau$ values between consecutive index
fluctuations and calculate the average return interval
$\overline{\tau}$.  In case (ii), we estimate average market behavior,
e.g., by analyzing all the 500 members of  the S\&P500 index. For
each $q$ and each company we calculate all $\tau_q$ values and their
average.

For any given value of $Q$ in order to improve statistics, we aggregate all the
$\tau$ values in one data set and calculate $\overline{\tau}$.  If the
pdf of large volatilities is asymptotically power-law distributed,
$P(|x|) \sim|x|^{-1-\alpha}$, and $P(|\tilde x|)\sim|\tilde
x|^{-1-\tilde\alpha}$, we propose a novel estimator which relates the
mean return intervals $\overline{\tau}_q$ with $\alpha$, where
$\overline{\tau}_q$ is calculated for both case (i) and case (ii).
Since on average there is one volatility above threshold $q$ for every
$\overline{\tau}_q$ volatilities, then
\begin{equation}     
1/\overline{\tau}_q\approx\int_q^\infty P(|x|)d|x|=P(|x|>q)\sim q^{-\alpha}.
\label{PR}
\end{equation}  
For both case (i) and case (ii), we calculate $\overline{\tau}_q$ for
varying $q$, and obtain an estimate for $\alpha$ through the
relationship
\begin{equation}     
\overline{\tau}_q\propto q^\alpha.
\label{BP}
\end{equation}   

We compare our estimate for $\alpha$ in the above procedure with the
$\alpha$ value obtained from $P(|R|>Q)$, using an alternative method of
Hill \cite{Hill}.  If the pdf follows a power law $P(x)\sim
Ax^{-(1+\alpha)}$, we estimate the power-law exponent $\alpha$ by
sorting the normalized returns by their size, $x_1>x_2>\ldots> x_N$,
with the result \cite{Hill}
\begin{equation}     
\alpha=(N-1)\left[\sum_{i=1}^{N-1}\ln\frac{x_i}{x_N}\right]^{-1},
\label{hills}
\end{equation}   
where $N-1$ is the number of tail data points. We employ the criterion
that $N$ does not exceed 10\% of the sample size which to a good extent
 ensures that the sample is restricted to the tail part of the pdf 
  \cite{Pagan96}.
 
A new method based on detrended covariance, detrended cross-correlations
analysis (DCCA), has recently been proposed \cite{PRL08}.  To quantify
power-law {\it cross-correlations\/} in non-stationary time series,
consider two long-range cross-correlated time series $\{y_i\}$ and
$\{y_i'\}$ of equal length $N$, and compute two integrated signals
$Y_k\equiv\sum_{i=1}^{k}y_i$ and $Y_k'\equiv\sum_{i=1}^{k}y'_i$, where
$k=1,\ldots,N$. We divide the entire time series into $N-n$ overlapping
boxes, each containing $n+1$ values. For both time series, in each box
that starts at $i$ and ends at $i+n$, define the ``local trend'' to be
the ordinate of a linear least-squares fit. We define the ``detrended
walk'' as the difference between the original walk and the local trend.

Next calculate the covariance of the residuals in each box $f^2_{\rm
  DCCA}(n,i)\equiv{1\over
  n-1}\sum_{k=i}^{i+n}(Y_k-Y_{k,i}')(Y_k-Y_{k,i}')$. Calculate the
detrended covariance by summing over all overlapping $N-n$ boxes of size
$n$,
\begin{equation}
\label{DXA}
   F_{\rm DCCA}^2(n) \equiv \sum_{i=1}^{N-n} f^2_{\rm DCCA} (n,i).
\end{equation}
If cross-correlations decay as a power law, the corresponding detrended
covariances are either always positive or always negative, and the
square root of the detrended covariance grows with time window $n$ as
\begin{equation}
F_{\rm DCCA}(n) \propto n^{\lambda_{\rm DCCA}},
\label{DCCAlambda}
\end{equation}
where $\lambda_{\rm DCCA}$ is the cross-correlation exponent.  If,
however, the detrended covariance oscillates around zero as a function
of the time scale $n$, there are no long-range cross-correlations.
  
When only one random walk is analyzed ($Y_k=Y_k'$), the {\it
  detrended covariance\/} $F^2_{\rm DCCA}(n)$ reduces to the {\it
  detrended variance} 
\begin{equation}
F_{\rm DFA}(n)\propto n^{\lambda_{\rm DFA}}
\end{equation}
used in the DFA method \cite{CKP}.

\section{Results of Analysis}

We first investigate the daily closing values of the S\&P500 index
adjusted for stock splits together with their trading volumes. 
In Fig.~1(a), we  show the cross-correlation function between
$|R_t|$ and $|\tilde R_t|$ and the cross-correlation function between
$R_t$ and $\tilde R_t$. The solid lines are 95\% confidence interval
for the autocorrelations of an i.i.d.  process.  The cross-correlation
function between $R_t$ and $\tilde R_t$ is practically negligible and
stays within the 95\% confidence interval. On the contrary, the
cross-correlation function between $|R_t|$ and $|\tilde R_t|$ is significantly
different than zero at the 5\% level  for more than 50 time lags.

 In
Fig.~1(b) we find, by using the DFA method \cite{CKP,DFA1}, that not
only $|R_t|$ \cite{Engle,Liu99}, but also $|\tilde R_t|$ exhibit 
power-law auto-correlations.  As an indicator that there is an
association between $|R_t|$ and $|\tilde R_t|$, we note that during
market crashes large changes in price are associated with large changes
in market volume. To confirm co-movement between $|R_t|$ and $|\tilde
R_t|$, in Fig.~1(b) we demonstrate that $|R_t|$ and $|\tilde R_t|$ are
power-law cross-correlated with the DCCA cross-correlation exponent (see
Methods section) close to the DFA exponent \cite{CKP,DFA1} corresponding
to $|R_t|$.  Thus, we find the cross-correlations between $|R_{t+n}|$
and $|\tilde R_t|$ not only at zero time scale $(n=0)$, but for a large
range of time scales.
  
\begin{figure}
\centering{\includegraphics[width=0.42\textwidth]{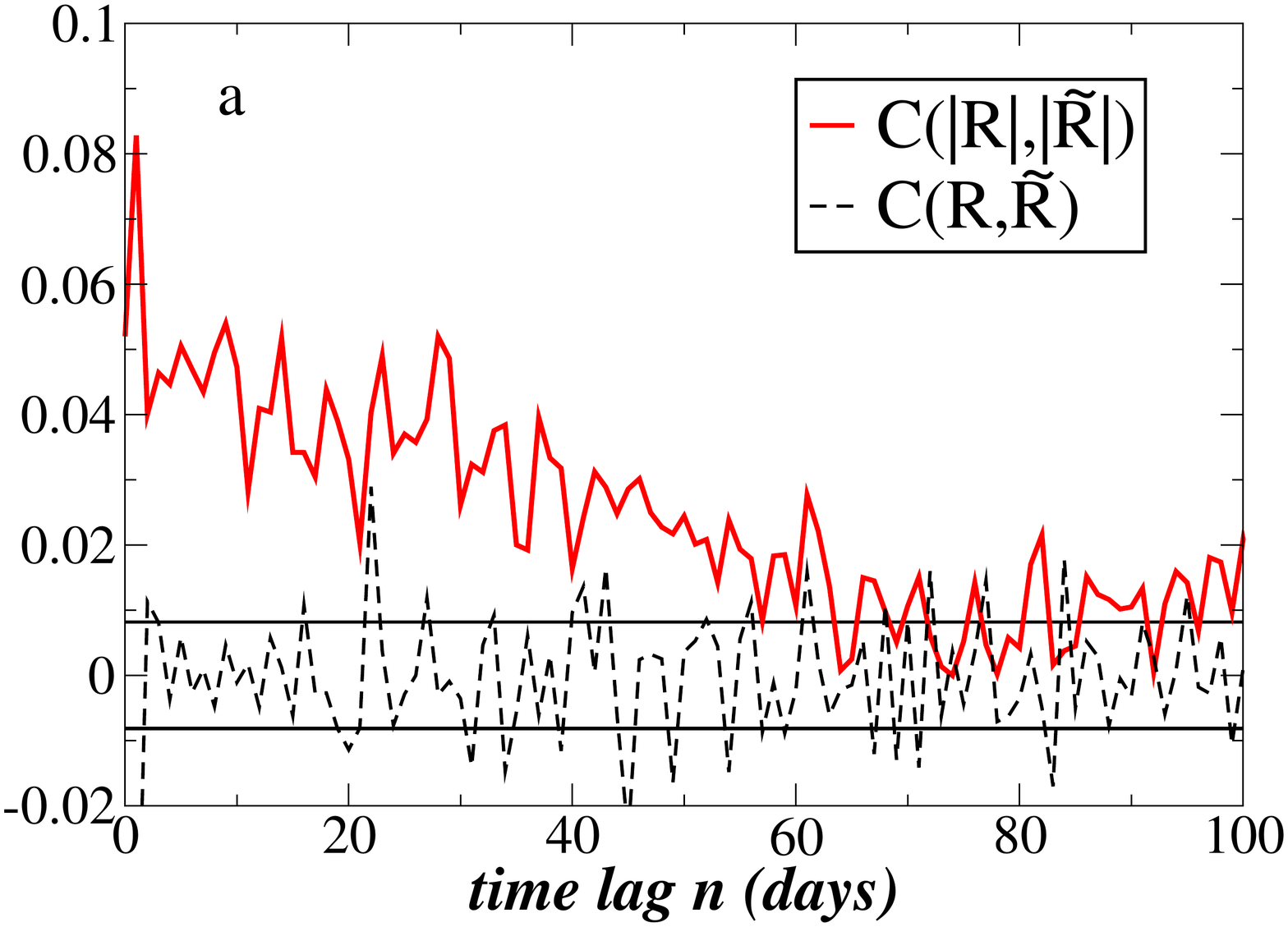}} 
\centering{\includegraphics[width=0.38\textwidth]{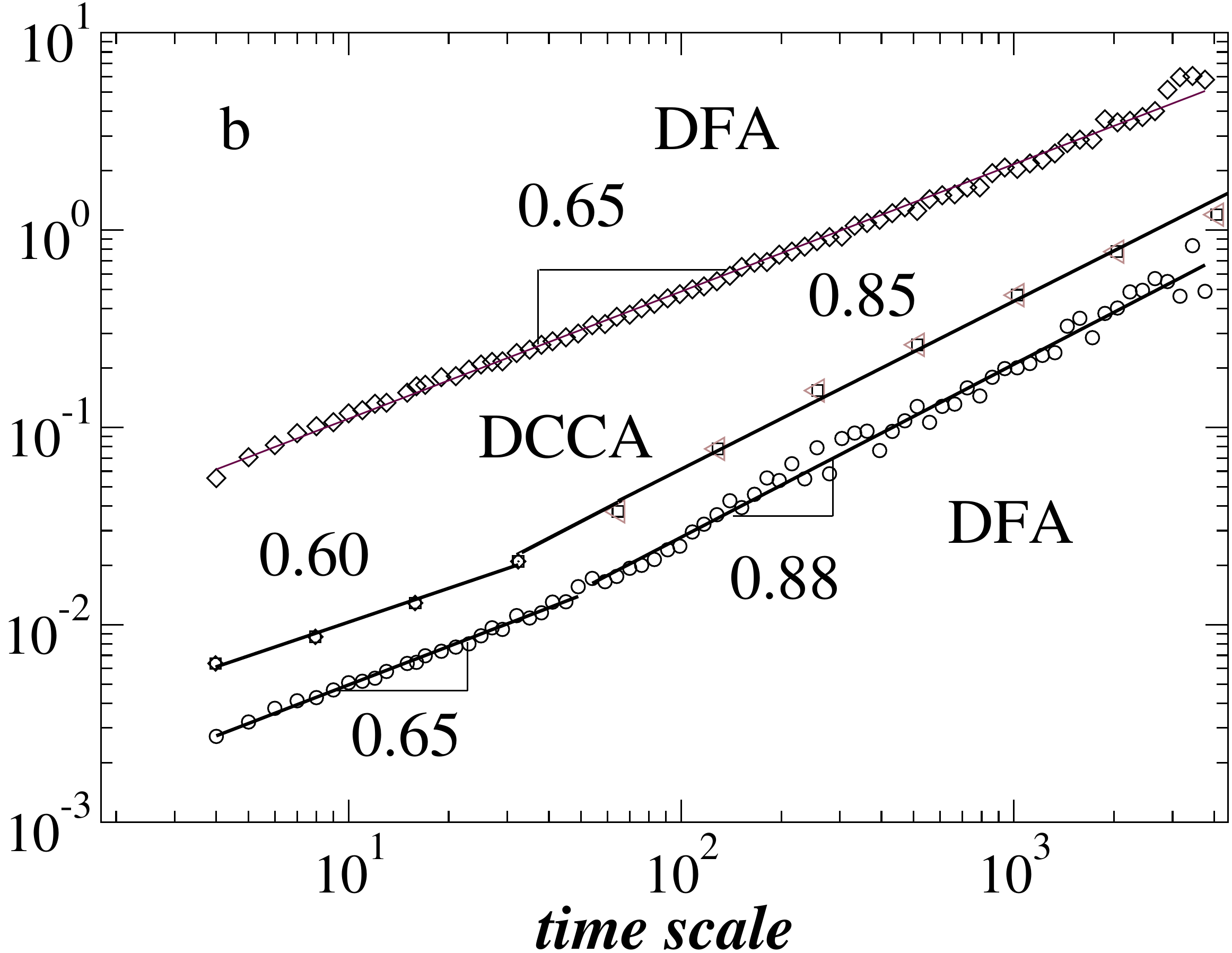}}
\caption{Auto-correlations and cross-correlations in absolute values of
  price changes $|R_t|$ of Eq.~(\ref{RS}) and trading-volume changes
  $|\tilde R_t|$ of Eq.~(\ref{RV}) for daily returns of the S\&P500
  index. (a) The cross-correlation function $C(R,\tilde R)$ between
  $R$ and $\tilde R$, and the cross-correlation function $C(|R|,|\tilde
  R|)$ between $|R|$ and $|\tilde R|$. 
  (b) For $R(t)$, and $\tilde R(t)$, we show the rms of the
  detrended variance $F_{\rm DFA}(n)$ for $|R|$ and $|\tilde R|$ and
  also the rms of the detrended covariance \cite{PRL08}, $F_{\rm
    DCCA}(n)$. The two DFA exponents $\lambda_{|R|}$ and
  $\lambda_{|\tilde R|}$ imply that power-law auto-correlations exist in
  both $|R|$ and $|\tilde R|$. The DCCA exponent implies the presence of
  power-law cross-correlations.  Power-law cross-correlations between
  $|R|$ and $|\tilde R|$ imply that current price changes depend upon
  previous changes, but also upon previous volume changes, and vice
  versa.  }
\label{fig.1}
\end{figure}

\begin{figure}
\centering{\includegraphics[width=0.38\textwidth]{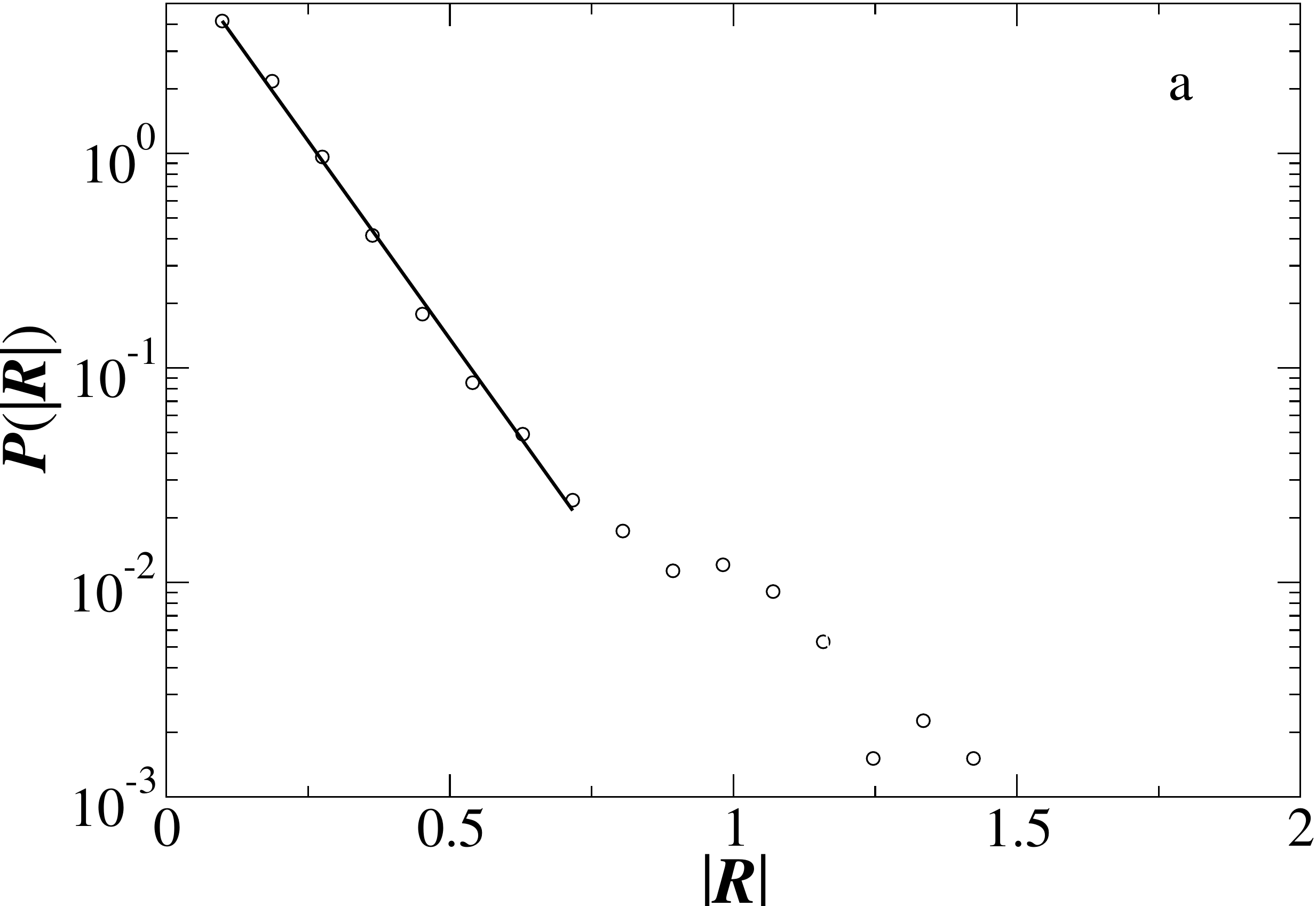}}
\centering{\includegraphics[width=0.46\textwidth]{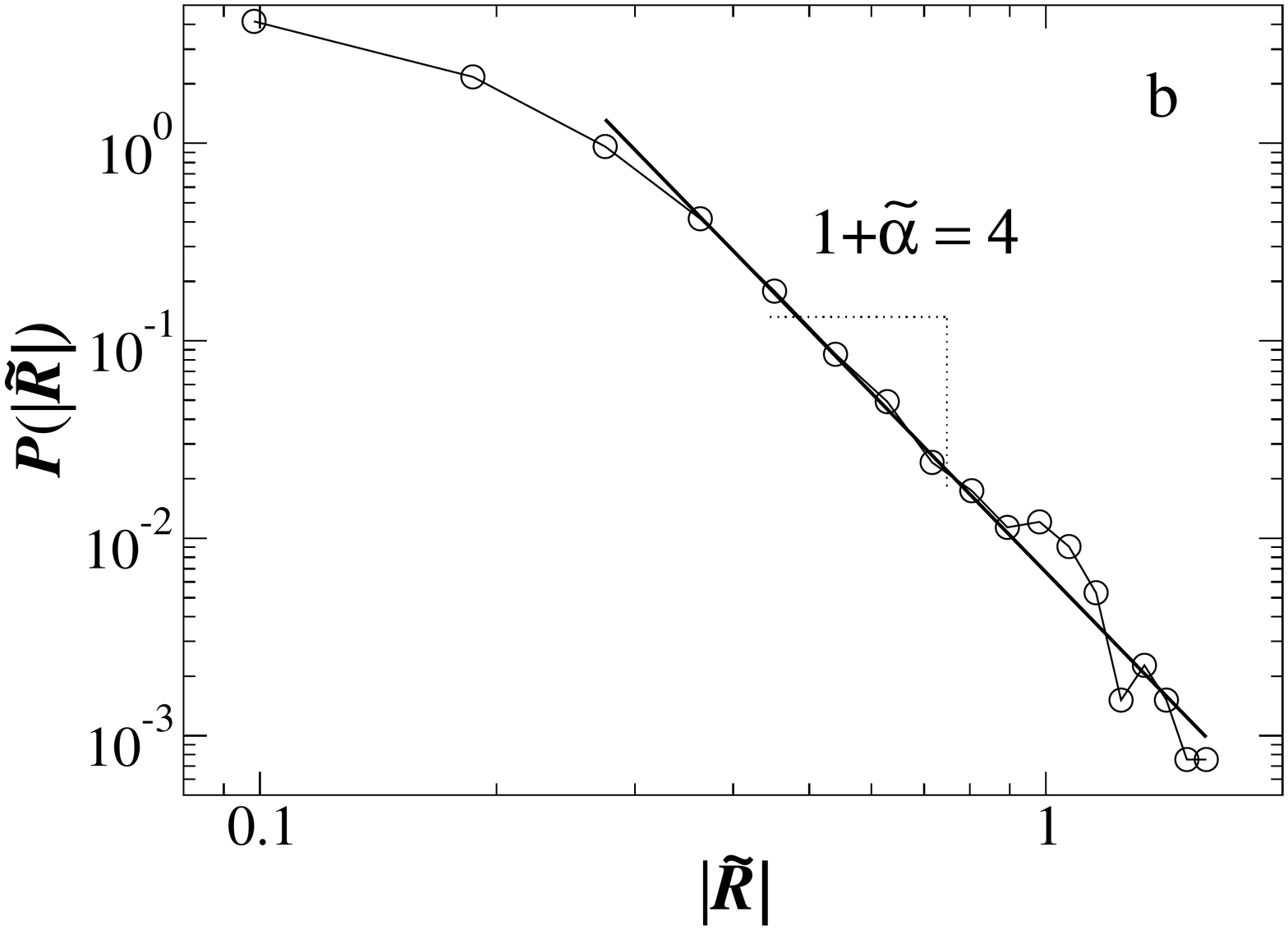}}
\centering{\includegraphics[width=0.38\textwidth]{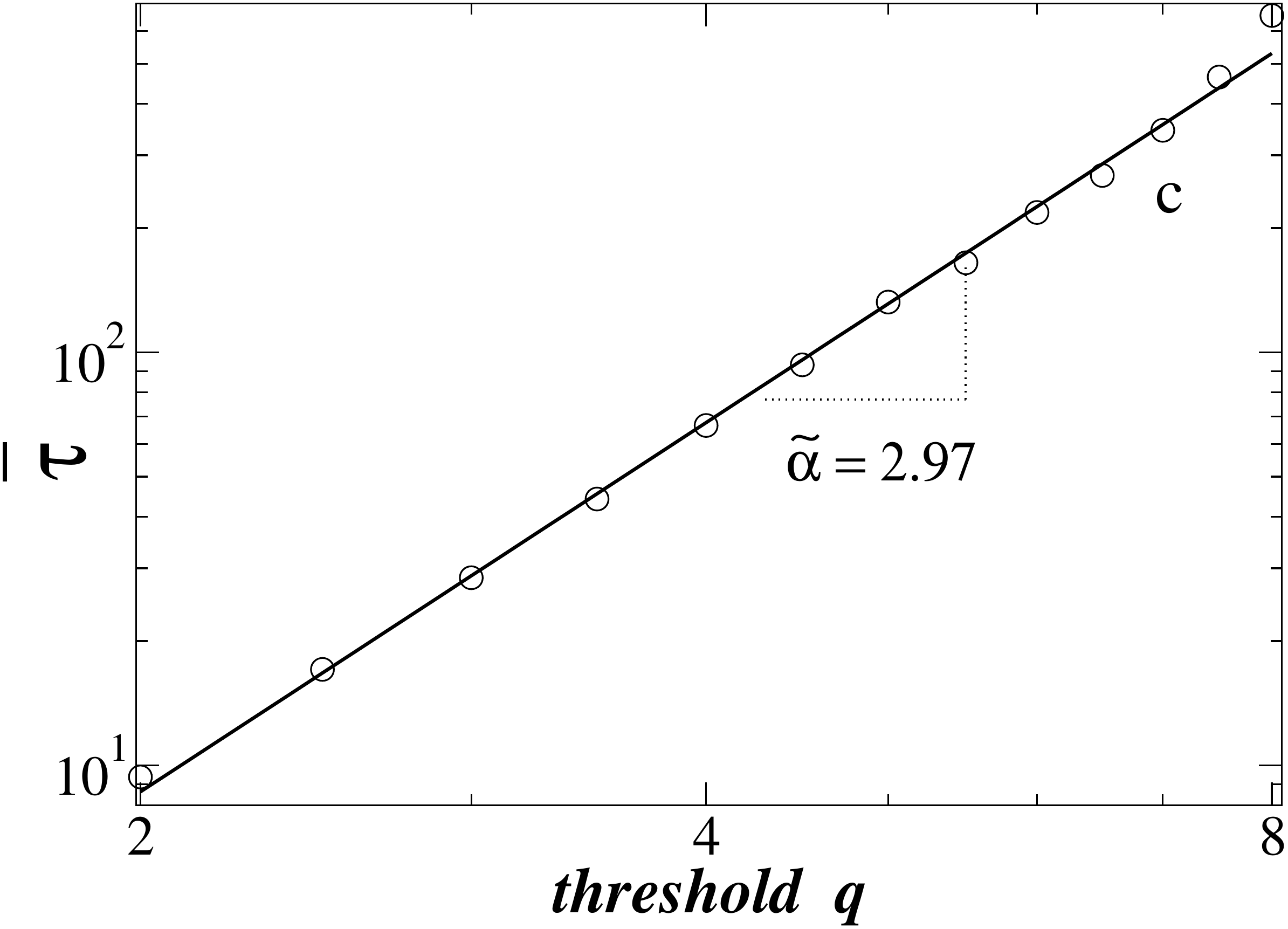}}
\caption{Pdf $P(|\tilde R|)$ of absolute value of differences in
  logarithm of trading volume, $\tilde R$, of Eq.~(\ref{RV}) for the
  S\&P500 index. (a) A log-liner plot $P(|\tilde R|)$. The solid line is
  an exponential fit. The tail part of pdf deviates from the fit in the
  central part. (b) Log-log plot of the pdf. The broad tail part can be
  explained by a power law $\tilde R^{1 + \tilde \alpha}$ with 
  $\tilde \alpha =3 \pm
  0.16$. (c) For the absolute values of changes in trading volume (see
  Eq.~\ref{RV}) the average return interval $\tau$ vs. threshold $q$ (in
  units of standard deviation $\sigma$) follows a power law, with
  exponent $\tilde \alpha = 2.97 \pm 0.02$.  The power law is consistent with
  inverse cubic law of the pdf.}
\label{fig.2}
\end{figure}

Having analyzed cross-correlations between corresponding (absolute)
changes in prices and volumes, we now investigate the pdf of the
absolute value of $\tilde R_t$ of Eq.~(\ref{RV}).  In order to test
whether exponential or power-law functional form fits better the data,
in Figs.~2(a) and (b) we show the pdf $P(\tilde R)$ in both linear-log
and log-log plot.  In Fig.~2(a) we see that the tail substantially
deviates from the central part of pdf which we fit by exponential
function.  In Fig.~2(b) we find that the tails of the pdf can be well
described by a power law $\tilde R^{1 +\tilde  \alpha}$ with exponent 
$\tilde \alpha=3\pm 0.16$, which supports an inverse cubic 
law---virtually the same as
found for average stock price returns \cite{Lux96,Gopi98,Gopi99}, and
individual companies \cite{Plerou99}.

In order to justify the previous finding, we employ two additional
methods.  First, we introduce a new method [described in Methods by
  Eqs.~(\ref{PR}) and (\ref{BP})] for a single financial index. We
analyze the probability that a trading volume change $\tilde R$ has an
absolute value larger than a given threshold, $q$. We analyze the time
series of the S\&P500 index for 14,922 data points.  First, we define
different thresholds, ranging from $2\sigma$ to $8\sigma$.  For each
$q$, we calculate the mean return interval, $\bar\tau$.  In Fig.~2(c) we
find that $q$ and $\bar\tau$ follow the power law of Eq.~(\ref{BP}),
where $\tilde \alpha = 2.97 \pm 0.02$.  We note that the better is the power
law relation between $\bar\tau_q$ and $q$ in Fig.~2(c), the better is
the power-law approximation $P(|\tilde R|>x) \approx x^{-\tilde  \alpha}$ for
the tail of the pdf $P(|\tilde R|)$.  In order to confirm our finding
that $P(|\tilde R|)$ follows a power law $P(|\tilde R|) \approx \tilde
R^{- \tilde \alpha -1}$ where $\tilde \alpha \approx 3$ obtained in 
Fig.~2(a) and
2(b), we also apply a third method, the Hill estimator \cite{Hill}, to a
single time series of the SP500 index.  We obtain $\tilde \alpha = 2.80 \pm
0.07$ consistent with the results in Fig.~2(a) and 2(b).

Next, by using the procedure described in case (ii) of {\it Methods\/},
we analyze 1,819 different time series of Eq.~(\ref{RV}), each
representing one of the 1,819 members of the NYSE Composite index.  For
each company, we calculate the normalized $|\tilde R_t|$ volatility of
trading volume changes of each company (see Eq.~(\ref{RVnorm})).  In
Figs.~3(a) and (b) we show the pdf in both linear-log and log-log plot.
In Fig.~3(a) we see that the broad central region of the pdf, from 2
$\sigma$ up to 15 $\sigma$, is fit by an exponential function. However,
the far tail deviates from the exponential fit.  In Fig.~3(b) we find
that the tails of the pdf from 15 $\sigma$ to up to 25 $\sigma$, are
described by a power law $\tilde R^{1 + \tilde \alpha}$ with exponent
$\tilde \alpha=4.65 \pm 1.00$.

\begin{figure}
\centering{\includegraphics[width=0.38\textwidth]{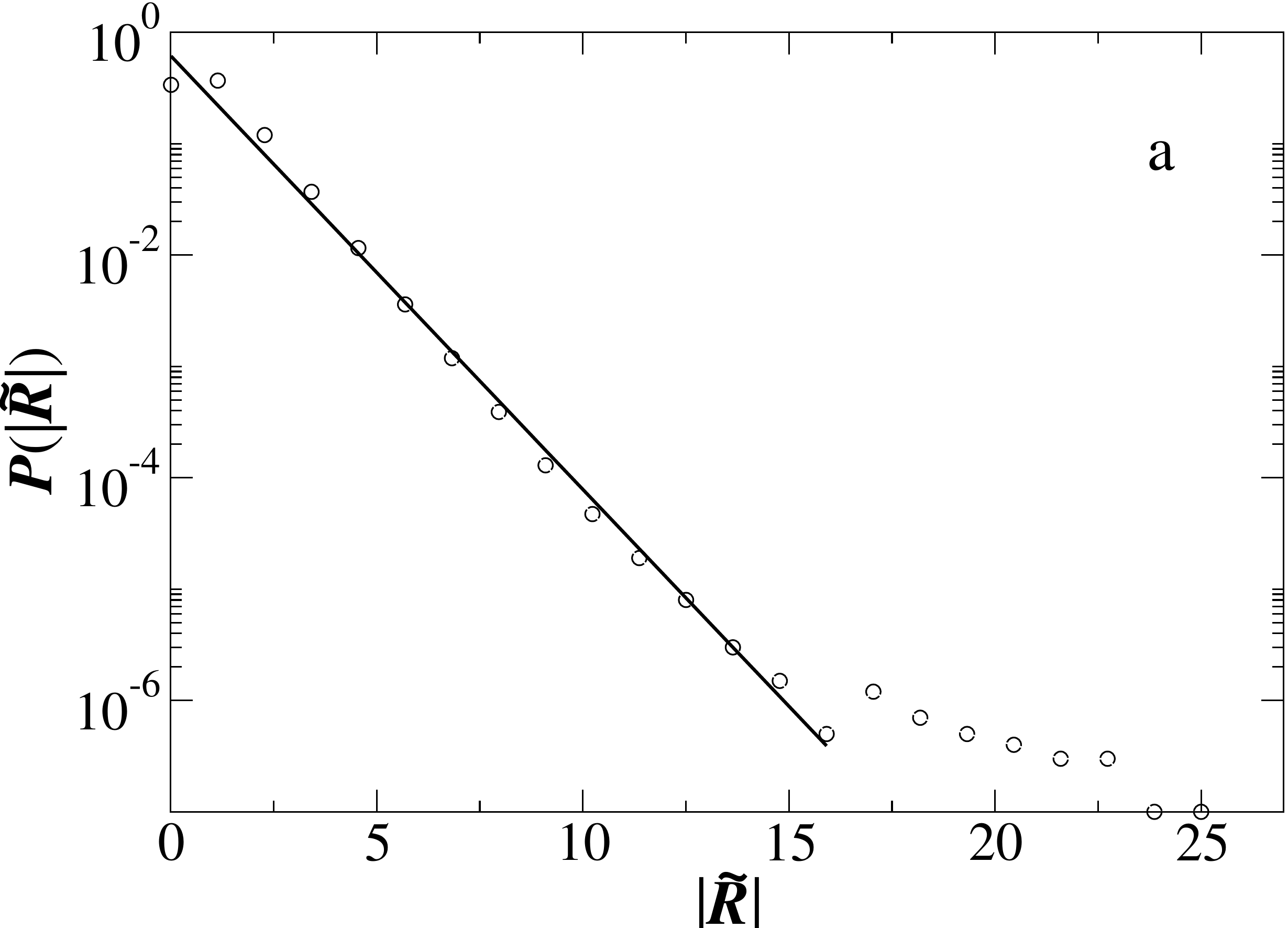}}
\centering{\includegraphics[width=0.38\textwidth]{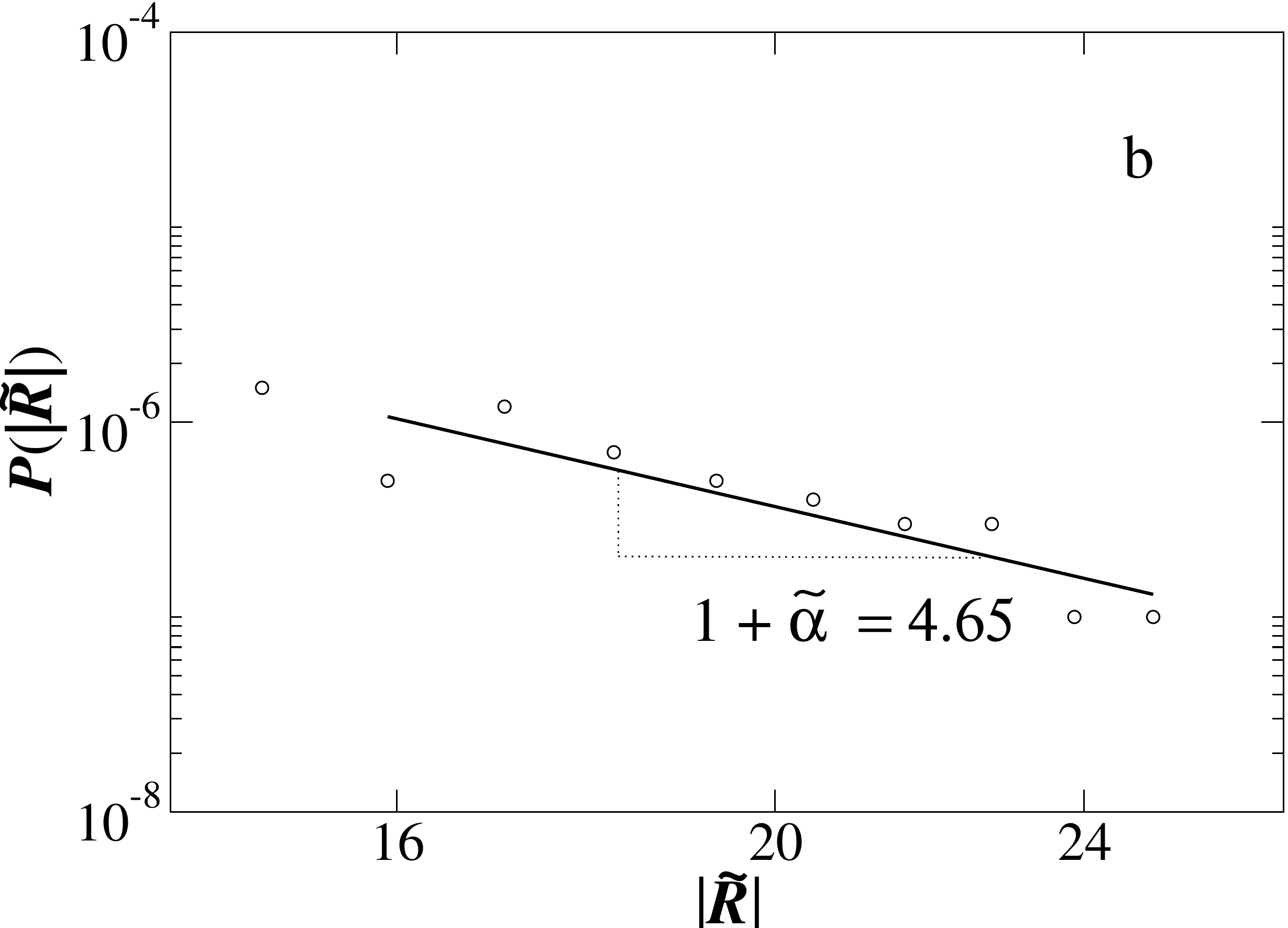}}
\centering{\includegraphics[width=0.38\textwidth]{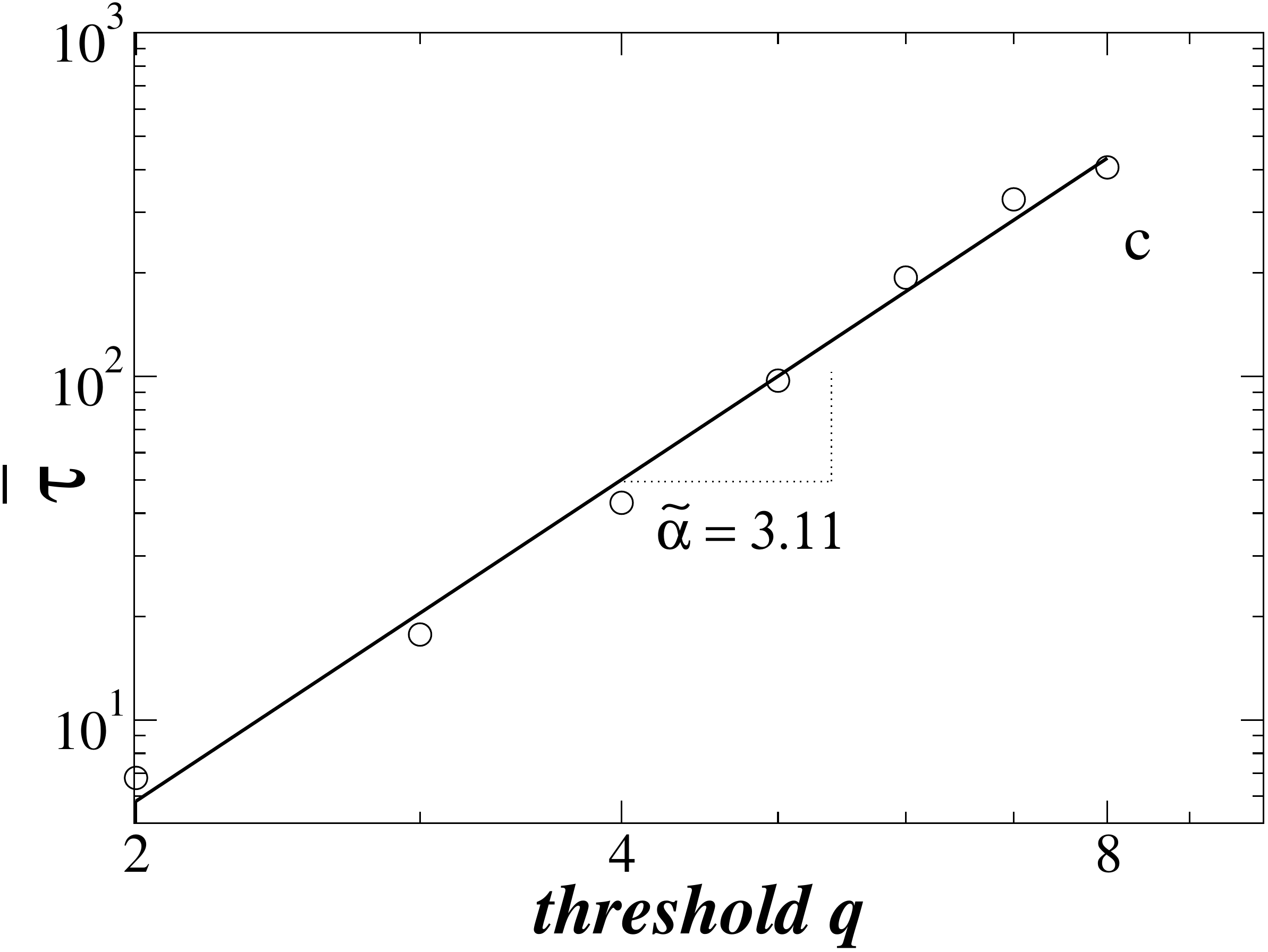}}
\caption{Pdf of absolute value of differences in logarithm of trading
  volume, $\tilde R$, of Eq.~(\ref{RV}) for the members of the NYSE
  Composite index.  We use the method described in the Methods
  section---case (ii)---for normalized volatilities of
  Eq.~(\ref{RVnorm}).  (a) From 1$\sigma$ to 15$\sigma$ we show the
  linear-log plot of the pdf $P(\tilde R)$. The straight line is
  exponential fit. The far tail of pdf deviates from the fit in the
  central region of pdf. (b) Log-log plot of pdf from 15$\sigma$ to
  25$\sigma$. The tail part of the pdf can be explained by a power law
  $\tilde R^{1 +\tilde  \alpha}$ with $\tilde \alpha=4.2\pm 0.26$.  
  (c) For the
  absolute values of changes in trading volume [see Eq.~(\ref{RV})] we
  show the average return interval $\bar\tau_q$ versus threshold $q$ (in
  units of a standard deviation). Up to 8$\sigma$, we show a power law
  with exponent $\tilde \alpha= 3.11 \pm 0.12$ which leads to the inverse
   cubic  law.}
\label{fig.3}
\end{figure}

Then, by employing the method described by Eqs.~(\ref{PR}) and
(\ref{BP}) we define different thresholds, $q$, ranging from $2\sigma$
to $8\sigma$ (different range than in Fig.~3(a)). We choose the lowest
$q$ equal to 2 since we employ the criterion that $N$ does not exceed
10\% of the sample size \cite{Pagan96}.  For each $q$, and each company,
we calculate the time series of return intervals, $\tau_q$.  For a given
$q$, we then collect all the $\tau$ values obtained from all companies
in one unique data set --- mimicking the market as a whole --- and
calculate the average return interval, $\bar\tau_q$.  In Fig.~3(c) we
find that $q$ and $\bar\tau_q$ follow an approximate inverse cubic law
of Eq.~(\ref{BP}), where $\tilde \alpha = 3.1 \pm 0.11$.  Our method is
sensitive to data insufficiency, so we show the results only up to 8
$\sigma$.  Clearly, this method gives the $\tilde \alpha$ value for the market
as a whole, not the $\tilde \alpha$ values for particular companies. By joining
all the normalized volatilities $|\hat{\tilde R}_t|$ obtained from 1,819
time series in one unique data set, we estimate Hill's exponent of
Eq.~(\ref{hills}), $\tilde \alpha = 2.82 \pm 0.003$, consistent with the 
value of exponent obtained using the method of Eqs.~(\ref{PR}) and (\ref{BP}).

 In the previous analysis we consider time series of 
 the companies comprising the NYSE Composite index of
 different lengths (from 10 to 11,966 data points). In order to prove that 
 the  Hill exponent of Eq.~(\ref{hills}) 
 is not affected  by the shortest time series, next we analyze
 only the time series longer than 3,000 data points (1,128
  firms in total). For the  Hill exponent  we obtain 
  $\tilde \alpha=2.81 \pm 0.003$,   that is the value practically the same
  as the one $(\tilde \alpha = 2.82 \pm 0.003)$  we obtained when short 
  time series were  considered as well.

We perform the method of Hill \cite{Hill}, and the method of
Eqs.~(\ref{PR}) and (\ref{BP}), also for the 500 members of the S\&P500
index comprising the index in July 2009.  There are in total $2,601,247$
data points for $\tilde R$ of Eq.~(\ref{RVnorm}).  For the thresholds,
$q$, ranging from $2\sigma$ to $10\sigma$, we find that $q$ and
$\overline{\tau}$ follow for this range an approximate inverse cubic law
of Eq.~(\ref{BP}), where $\tilde \alpha = 3.1 \pm 0.12$.  We estimate the Hill
exponent of Eq.~(\ref{hills}) to be $\tilde \alpha=2.86\pm 0.005$, with the
lowest $Q=2$.
    
In order to find what is the functional form for trading-volume changes
at the world level, we analyze 28 worldwide financial indices using the
procedure described in {\it Methods\/} [case(ii)].  For each $q$, and
for each of the 28 indices, we calculate the values for the return
interval $\tau$.  Then for a given $q$, we collect all the $\tau$ values
obtained for all indices and calculate the average return interval
$\bar\tau_q$.  In Fig.~4(a), we find a functional dependence between $q$
and $\overline{\tau}$ which can be approximated by a power law with
exponent $\tilde \alpha = 2.41 \pm 0.06$. We also calculate $\overline{\tau}$
 vs. $q$ for different levels of financial aggregation. 
 
Finally, in addition to trading-volume changes, we employ for stock
price changes our procedure for identifying power-law behavior in the
pdf tails described in {\it Methods\/} [case (ii)].  The pdf of stock
price changes, calculated for an ``average'' stock, is believed to
follow $P(R)\approx R^{-(1+\alpha)}$ where $\alpha\approx 3$, as
empirically found for wide range of different stock markets
\cite{Lux96,Gopi99}.

Next we test whether this law holds more generally. To this end, we
analyze the absolute values of price changes, $|R_t|$ [see
  Eq.~(\ref{RS})], for five different levels of financial aggregation:
(i) Europe, (ii) Asia, (iii) North and South America, (iv) the world
without the USA, and (v) the entire world. For each level of
aggregation, we find that the average return interval $\bar\tau_q\sim
q^{-3}$.

\begin{figure}
\centering{\includegraphics[width=0.4\textwidth]{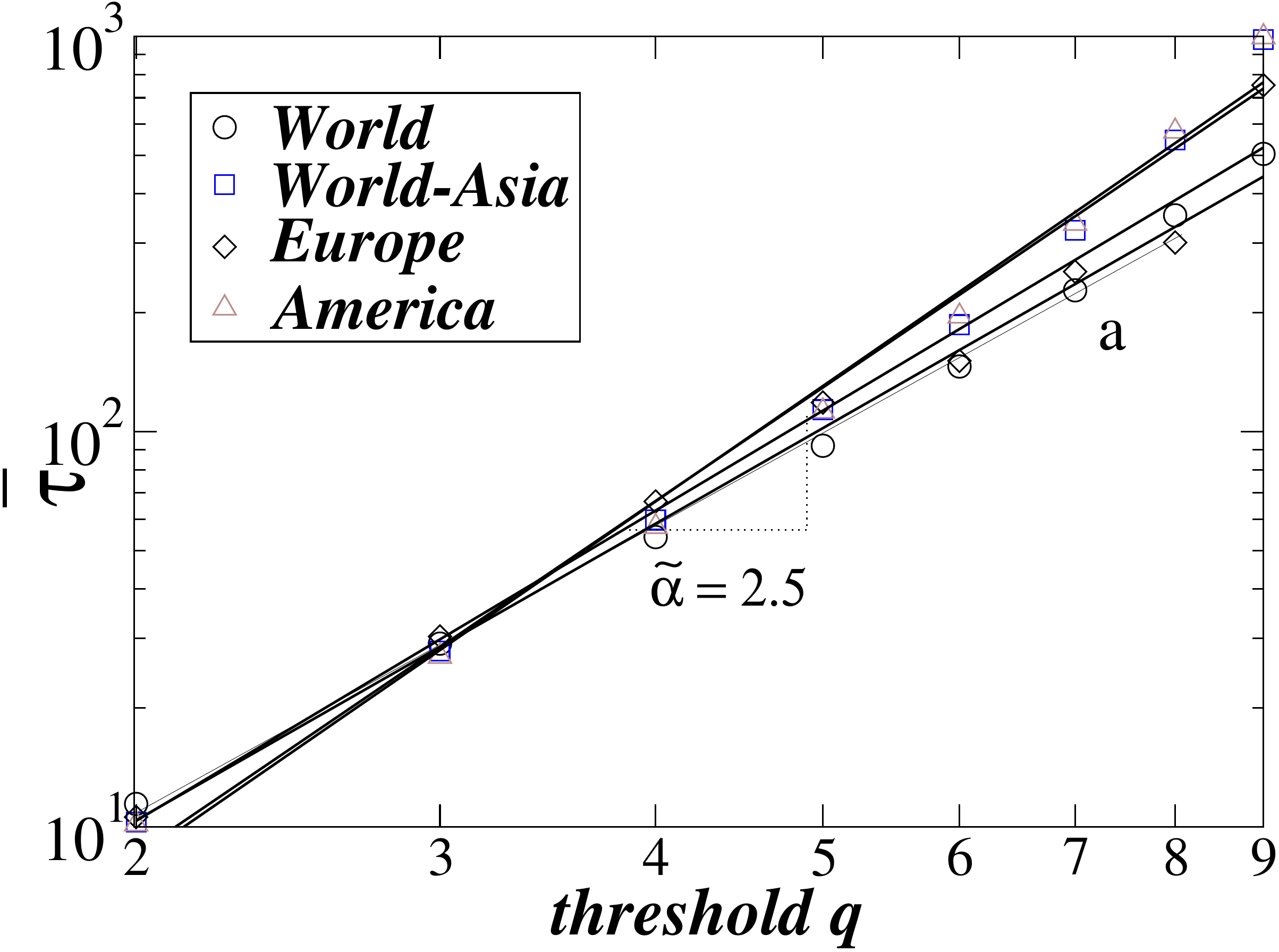}}
\centering{\includegraphics[width=0.4\textwidth]{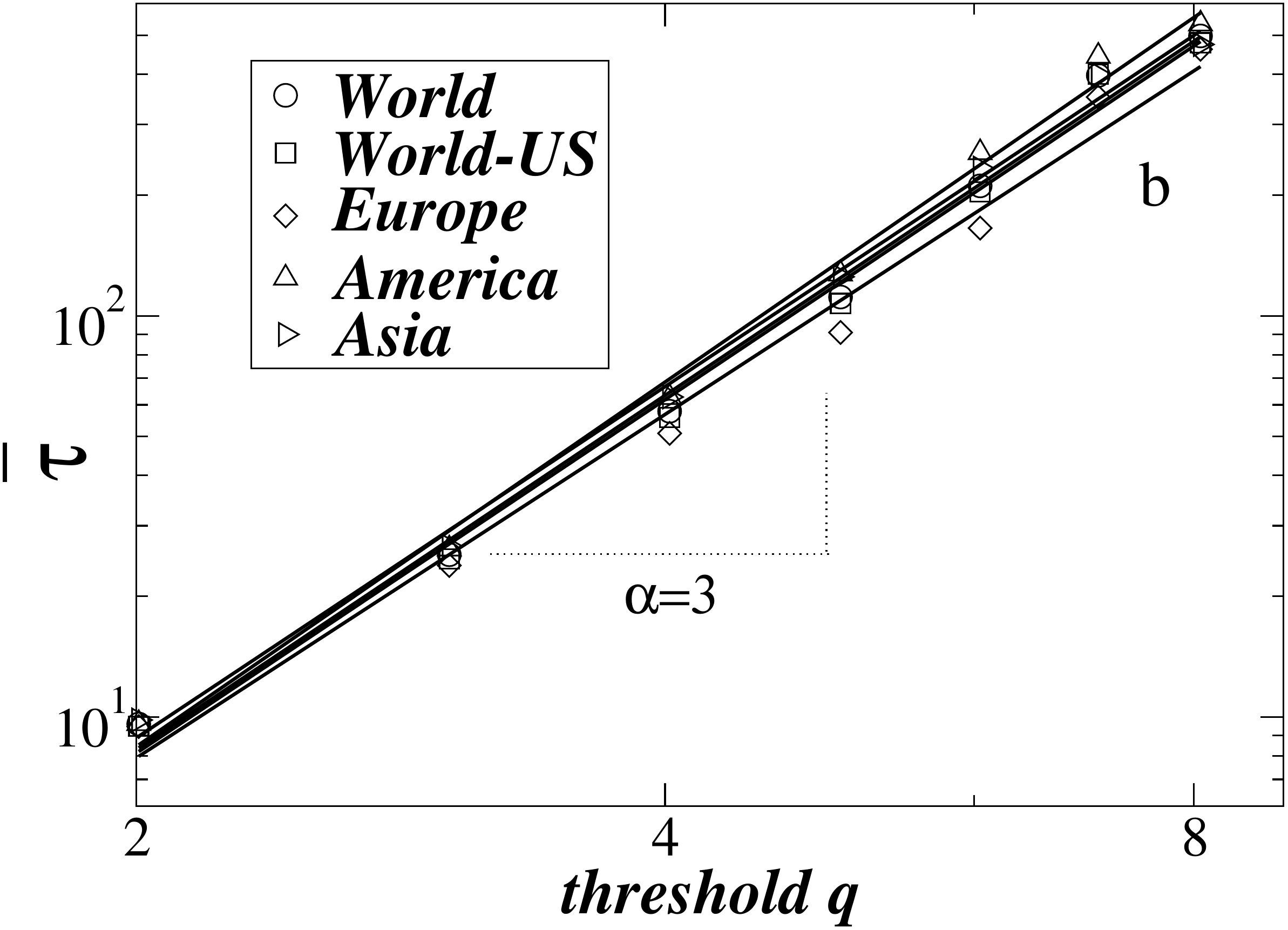}}
\caption{Power-law correlations for world-wide financial indices in (a)
  absolute values of price changes $(|\tilde R|)$ and (b) absolute
  values of trading-volume changes $(|R|)$.  We use the method described
  by Eqs.~(7)--(8).  (a) The average return interval $\bar\tau$
  vs. threshold $q$ (in units of standard deviation) for absolute values
  of trading-volume changes.  For each of 28 worldwide financial
  indices, we calculate the corresponding $\bar\tau_q$ values. Then we
  collect all the $\bar\tau$ values obtained from different indices, and
  show $\bar\tau_q$ versus $q$. Up to 8 standard deviations, we find a
  power law with exponent $\tilde \alpha = 2.41 \pm 0.06$.  (b) The average
  return interval $\bar\tau_q$ vs. threshold $q$ for absolute values of
  price changes [see Eq.~(\ref{RS})] for different levels of
  aggregation.  For each of five different types of aggregation
  reported, we find that $\bar\tau$ versus $q$ exhibits a power law with
  an exponent very close to $\alpha=3$.}
\end{figure}

\section{Model}

In order to model long-range cross-correlations between $|R_t|$ and
$|\tilde R_t|$, we introduce a new joint process for price changes
\begin{equation}
   \epsilon_t =   \sigma_t \eta_{t}
   \label{eps1}
\end{equation}
\begin{equation}
   \sigma^2_t = \omega +
   \alpha~\epsilon_{t-1}^2  +  \beta~\sigma^2_{t-1} + 
  \tilde   \gamma ~   \tilde  \epsilon^{2}_{t-1}
 \label{sht}
\end{equation}
and for trading-volume changes
\begin{equation}
  \tilde \epsilon_t =    \tilde \sigma_t \tilde  \eta_{t}
\end{equation}
\begin{equation}
  \tilde \sigma^{2}_t =  \tilde \omega +
  \tilde  \alpha ~  \tilde \epsilon^{2}_{t-1}  +  
 \tilde   \beta ~ \tilde \sigma^{2}_{t-1} +  \gamma ~\epsilon^{2}_{t-1}.
 \label{vht}
\end{equation}
If $\gamma=\tilde\gamma=0$, Eqs.~(\ref{eps1})--(\ref{vht}) reduce to two
separate processes of Ref.~\cite{Boll}. Here $\eta_{t}$ and
$\tilde\eta_t$ are two i.i.d. stochastic processes each chosen as
Gaussian distribution with zero mean and unit variance. In order to fit
two time series, we define free parameters $\omega$, $\alpha$, $\beta$,
$\gamma$, $\tilde\omega$, $\tilde\alpha$, $\tilde\beta$, $\tilde\gamma$,
which we assume to be positive \cite{Boll}.  The process of
Eqs.~(\ref{eps1})--(\ref{vht}) is based on the generalized
autoregressive conditional heteroscedasticity (GARCH) process (obtained
from Eqs.~(\ref{eps1})-(\ref{sht}) when $\tilde \gamma =0$) introduced to
simulate long-range auto-correlations through $\beta \ne 0$. The GARCH
process also generates the power-law tails as often found in empirical
data [see \cite{Lux96,Gopi98,Gopi99,Plerou99}, and also Fig.~2(b)].  In
the process of Eqs.~(\ref{eps1})--(\ref{vht}) we obtain
cross-correlations since time-dependent standard deviation $\sigma_t$
for price changes depends not only on its past values (through $\alpha$
and $\beta$), but also on past values of trading-volume errors
($\tilde\gamma$). Similarly, $\tilde\sigma_t$ for trading-volume changes
depends not only on its past values (through $\tilde\alpha$ and $\tilde
\beta$), but also on past values of price errors ($\gamma$).

For the joint stochastic process of Eqs.~(\ref{eps1})-(\ref{vht}) with
$\beta=\tilde\beta=0.65$, $\alpha=\tilde\alpha=0.14$,
$\gamma=\tilde\gamma=0.2$, we show in Fig.~5(a) the cross-correlated
time series of Eqs.~(\ref{sht}) and (\ref{vht}). In Fig.~5(b) we show
the auto-correlation function for $|\epsilon_t|$ and the cross-correlation
function which practically overlap due to the choice of parameters.
 
If stationarity is assumed, we calculate the expectation of
Eq.~(\ref{sht}) and (\ref{vht}) and since, e.g., ${\rm
  E}(\sigma^2_t)={\rm E}(\sigma^2_{t-1})={\rm
  E}(\epsilon_{t-1}^2)=\sigma_0^2$, we obtain
$\sigma_0^2(1-\alpha-\beta)=\omega+  \tilde \gamma \tilde \sigma_0^{2}$, and similarly
$\tilde \sigma_0^{2}(1-\tilde \alpha - \tilde\beta)=
 \tilde \omega + \gamma \sigma_0^2$. So,
stationarity generally assumes that $\alpha+\beta<1$ as found for the
GARCH process \cite{Boll}.  However, for the choice of parameters in the
previous paragraph for which $\sigma_0=\tilde\sigma_0$ stationarity
assumes that $\sigma_0^2(1-\alpha-\beta- \tilde \gamma)=\omega$.  This result
explains why the persistence of variance measured by $\alpha+\beta$
should become negligible in the presence of volume in the GARCH process
\cite{Lamo90}. In order to have finite $\sigma_0^2$, we must assume
$\alpha+\beta+ \tilde \gamma  < 1$.  

 It is also possible to consider IGARCH and FIGARCH  processes  with
  joint  processes for price and volume change, a potential avenue for future
   research \cite{Boll2}. 
   
\begin{figure}
\centering{\includegraphics[width=0.5\textwidth]{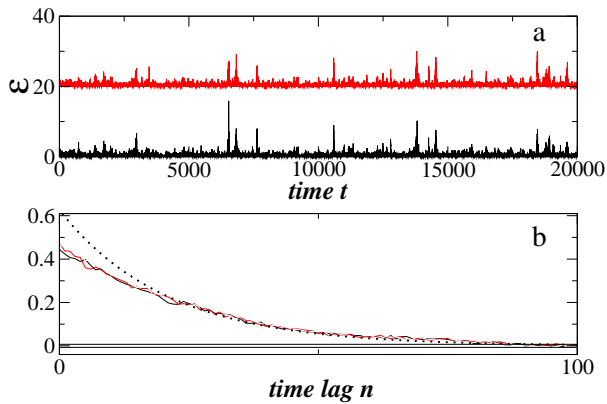}}
\caption{Cross-correlations between two time series generated from the stochastic
  process of Eqs.~[\ref{eps1}-\ref{vht}], with $\beta= \tilde \beta
  =0.65$, $\alpha= \tilde \alpha=0.14$, $\gamma= \tilde \gamma =0.2$,
  and $\omega= \tilde \omega =0.01$. In panel (a) we show the 
  time series $\epsilon$ and $\tilde\epsilon$ of Eqs.~[\ref{eps1}-\ref{vht}], where the latter time
  series is shifted for clarity. These two time series follow each other due
  to the  terms $\gamma \ne 0 $ and $ \tilde \gamma \ne 0$.  
  In panel (b) we show the
  auto-correlation function $A(n)$ for $|\epsilon_t|$ and 
 the cross-correlation function $C(|\tilde\epsilon|,|\epsilon|) $. 
  The 95\% confidence intervals for no cross-correlations are shown 
  (solid lines)
   along with the best exponential fit of $A(n)$ (dotted curve). 
}
\end{figure}

\section{Summary}

In order to investigate possible relations between price changes and 
 volume changes, we analyze the properties of $|\tilde R|$, the logarithmic volume change. We hypothesize that  the underlying processes
 for  logarithmic price  change $|R|$ and   logarithmic volume change  $|\tilde R|$  are  similar. Consequently, 
 we use the traditional methods that are used to analyze changes in trading price to analyze changes in trading volume. Two major
  empirical  findings are: 
 
(i) we analyze a well-known U.S. financial index, 
 the S\&P500 index  over the
59-year period 1950-2009, and find 
  power-law 
 cross-correlations between $| \tilde R|$ and $|R|$. We  find 
 no  cross-correlations between $\tilde R$ and $R$. 
 
(ii) we demonstrate that, at different levels of aggregation, ranging from 
 the S\&P500 index, to aggregation of
different world-wide financial indices,  $|\tilde R|$
approximately follows the same cubic law as  $|R|$.  
Also, we find that the central region 
 of the pdf, $P(|\tilde R|)$, follows an exponential function as reported for  {\it annually} recorded variables, 
 such  as GDP \cite{Lee,PRE08},   company sales \cite{Michael}, 
 and stock prices  \cite{EPL09}.

In addition to empirical findings, we
 offer two theoretical  results: 
 
(i) to estimate the tail exponent $\tilde \alpha$ for the pdf of $|\tilde R|$, 
we develop an estimator which
relates $\tilde \alpha$ of the 
cdf $P(|\tilde R|> x ) \approx x^{- \tilde\alpha}$ to the average return
interval $\overline{\tau}_q$ between two consecutive volatilities 
above a threshold $q$ \cite{Yama05}.  

(ii) we  introduce a
  joint stochastic  process for modeling simultaneously 
     $|R|$ and $|\tilde R|$, which 
   generates  the cross-correlations between  $|R|$ and $|\tilde R|$. We also
    provide  conditions for stationarity.    


\begin{acknowledgments}

We thank NSF and the Ministry of Science of Croatia  for financial support.

\end{acknowledgments}

\end{document}